\documentclass[11pt,aps,nofootinbib,prd,aps,epsf,floats,axodraw,
               amsmath,amssymb,amsfonts]{revtex4}
\usepackage{amsmath, amssymb}
\newcommand{\mathsym}[1]{{}}

\usepackage{graphicx}% Include figure files
\usepackage{amsmath}
\usepackage{amssymb}
\usepackage{bm}% bold math
\newcommand{\be}{\begin{equation}}
\newcommand{\ee}{\end{equation}}
\newcommand{\bea}{\begin{eqnarray}}
\newcommand{\eea}{\end{eqnarray}}

\newcommand{\rem}[1]{}
\newsavebox{\PSLASH}
 \sbox{\PSLASH}{$p$\hspace{-1.8mm}/}
 
\renewcommand{\theequation}{\thesection.\arabic{equation}}
\newcounter{saveeqn}
\newcommand{\add}{\addtocounter{equation}{1}}
\newcommand{\alpheqn}{\setcounter{saveeqn}{\value{equation}}%
\setcounter{equation}{0}%
\renewcommand{\theequation}{\mbox{\thesection.\arabic{saveeqn}{\alph{equation}}}}}
\newcommand{\reseteqn}{\setcounter{equation}{\value{saveeqn}}%
\renewcommand{\theequation}{\thesection.\arabic{equation}}}

 \newsavebox{\notrightarrow}
 \sbox{\notrightarrow}{$\to$\hspace{-4mm}/}
 
 \newsavebox{\PARTIALSLASH}
 \sbox{\PARTIALSLASH}{$\partial$\hspace{-1.6mm}/}
 
 \newsavebox{\ASLASH}
 \sbox{\ASLASH}{$A$\hspace{-2.1mm}/}
 
 \newsavebox{\KSLASH}
 \sbox{\KSLASH}{$k$\hspace{-1.8mm}/}
 
 \newsavebox{\LSLASH}
 \sbox{\LSLASH}{$\ell$\hspace{-1.8mm}/}
 
 \newsavebox{\QSLASH}
 \sbox{\QSLASH}{$q$\hspace{-1.8mm}/}
 
 \newsavebox{\DSLASH}
 \sbox{\DSLASH}{$D$\hspace{-2.2mm}/}
 
 \newsavebox{\DbfSLASH}
 \sbox{\DbfSLASH}{${\mathbf D}$\hspace{-2.8mm}/}
 
 \newsavebox{\DELVECRIGHT}
 \sbox{\DELVECRIGHT}{$\stackrel{\rightarrow}{\partial}$}
 
 \newcommand{\blue}{\IfColor{\textCadetBlue}{}}
\newcommand{\black}{\IfColor{\textBlack}{}}
\newcommand{\red}{\IfColor{\textRed}{}}
\newcommand{\green}{\IfColor{\textOliveGreen}{}}
\newcommand{\lila}{\IfColor{\textRedViolet}{}}

\begin{document}
\begin{flushright}
 [math-ph]
\end{flushright}
\title{Axial Symmetry, Anti-BRST Invariance, and Modified Anomalies}

\author{Amir Abbass Varshovi}\email{amirabbassv@ipm.ir}

\affiliation{
   School of Mathematics, Institute for Research in Fundamental Sciences (IPM).\\
   School of Physics, Institute for Research in Fundamental Sciences (IPM).\\
                                     Tehran-IRAN}
\begin{abstract}
       \textbf{Abstract\textbf{:}} It is shown that anti-BRST symmetry is the quantized counterpart of local axial symmetry in gauge theories. An extended form of descent equations is worked out which yields a set of modified consistent anomalies.
\end{abstract}
 \maketitle
%%%%%%%%%%%%%%%%%%%%%%%%
\section{Introduction}\label{introduction}
\indent There are intimate correlations between physical objects in the quantum theory of gauge fields, such as anomalies, (anti-) BRST symmetry, (anti-) ghosts, ..., and geometric-topological concepts in differential geometry, including Chern characters, index theorem, principal bundles, and topological invariants, which have been extensively studied in last few decades with spectacular results and vast amount of applications [1]. Specially, anomaly of axial symmetry at quantum levels was one of the first significant gates toward illustration of such topological features inside a physical gauge theory, the quantum electro-dynamics, as an Abelian model for gauge theories.
 \par Historically, soon after the pioneering articles of Adler, Bell and Jackiw [2, 3] on the anomalous axial Ward identity for $U(1)$-gauge theory, Bardeen [4] showed that for more general theories a generalized version of anomalous term exists. More precisely, Bardeen proved that the root of the anomalous behavior is not entirely confined in axial currents, but in fact, it is basically hidden in the renormalization of the theory. In the other words, the anomalous terms come from those Feynman diagrams with few enough vertices which diverge linearly at the least. But, however, despite of this primitive topological dependence, the critical topological meaning of anomalous Ward identity was still not clear. Bardeen also showed that the scalar and the pseudo-scalar fields do not appear in the anomalous terms after renormalizing the theory, and therefore, the anomaly merely depends upon the gauge fields, or more precisely upon the curvature tensor, provided by gauge invariance of observables in gauge theories. It was also known that the whole anomaly can never completely be removed by using different forms of counter terms, the fact of which could be reported as an evidence for topological structure of anomalous Ward identities.
\par The consistency condition discovered by Wess and Zumino [5] was essentially an alternative criterion for gauge invariance of anomalous terms. The Bardeen's results were also confirmed by the consistency condition [5], which was then translated to topological structure for consistent anomalies as a non-trivial cohomology class of gauge transformation group.
\par Fujikawa was essentially the first one who shed the light on relationship between anomalies and topology [6-9]. He precisely proved that the anomalies are basically rooted in the topology of principal bundle of which the gauge theory is defined over. Specially, Fujikawa showed that the reason of axial anomaly lies in non-equality of dimensions of kernels of left- and right-handed Dirac operators as two mutually adjoint differential operators. This non-equality could be reported as breaking the classical axial symmetry at quantum levels, appearing in variation of the path integral measure of fermionic partition function against local axial transformations [6]. Indeed, Fujikawa rigorously proved that anomalies are intimately related to the celebrated Atiyah-Singer index theorem, which basically counts the analytic index of an elliptic pseudo-differential operator over a vector bundle (such as the Dirac operator on the spinor bundle) with integrating a specific topological differential form, a characteristic class, over the base manifold, substantiating an equivalence between analytic and topological structures [10, 11]. Particularly, Fujikawa demonstrated that the $n$th Chern class of $2n$-dimensional base manifold, the space-time, describes the topological structure of axial anomalous Ward identity through the index theorem. By this a fascinating feature for anomalous behaviors of gauge theories was worked out via spectacular achievements of modern differential geometry.
 \par On the other hand, the Faddeev-Popov path integral quantization method [12] for non-Abelian gauge theories led to an alternative formulation of classical gauge symmetry, the BRST symmetry, which was basically reflected from differential structures of the connection space over the principal bundle of the gauge theory [13, 14]. Indeed, the Faddeev-Popov quantization method replaced the classical gauge symmetry by its quantized version, the BRST invariance. In this formulation two kind of unphysical fields, ghost and anti-ghost, emerged in the quantum Lagrangian within the path integral setting together with Fermi-Dirac statistics and scalar behaviors (e.g.; spinless fermions).
 \par Although there was no expectation to see more symmetries after quantizing the gauge theories, but anti-BRST, another quantized symmetry of quantum Lagrangian [15, 16], emerged through the Faddeev-Popov quantization approach without any classical counterpart. It was also shown that a deep duality exists between BRST and anti-BRST symmetries and respectively for ghost and anti-ghost fields [17], but the precise correlation was still unclear geometrically.
\par After Stora and Zumino [18-20] discovered a sequence of descent equations which could relate the consistent anomalies and consistent Schwinger terms [21] through a set of deRham-BRST differential equations, Faddeev showed that this new point of view could be explained in the setting of cohomology algebras [22], and in precise, of the cohomology of gauge group, which was proven with details by Zumino [23]. Then, some rigorous geometric models for Faddeev-Popov quantization method, ghosts, BRST transformations and BRST cohomology were constructed to explain the path integral quantization of gauge theories in the setting of differential geometry [24-32] (for more collected and recent works see [33-36]). These geometric formulations led to elegant topological descriptions of consistent anomalies and consistent Schwinger terms [37-39]. In this geometric point of view, ghosts are left invariant operator valued 1-forms over the gauge transformation group and can be described by principal connections over the principal bundle of the moduli space of connections.
 \par The geometric procedure produces an alternative setting for Faddeev-Popov path integral quantization method. Indeed, this formulation can be considered as the geometric version of BRST quantization of gauge theories [40-44]. More precisely, although the structure of BRST invariance made its first appearance as an accidental symmetry of the quantum action, but it became apparent that it takes its most sophisticated setting in the Hamiltonian formulation of gauge theories, where it is used in the homological reduction of Poisson algebra of smooth functions on a symplectic manifold on which a set of first class constraints is defined. This constructive formulation lead to more drastic and deeper understanding of ghost fields, gauge fixing, and eventually BRST transformations in terms of symplectic geometry. Such geometric method of quantization of gauge theories based on symplectic data is conventionally referred to as geometric BRST quantization [45-47].
 \par In this article the main geometric ideas of BRST quantization is applied for studying some general version of gauge theories which admit the local axial symmetry. It is shown that because of enlargement of the gauge transformation group, the algebra of ghosts and BRST transformations get larger. Then, it is seen that the behaviors of new geometric objects are in complete agreement with those of anti-ghosts and anti-BRST transformations in the standard fashions [15, 16]. Indeed, a generalized version of BRST and anti-BRST transformations is found by considering the local axial symmetry. Strictly speaking, this consideration leads to an enlargement of the algebra generated by gauge fields, ghosts and anti-ghosts, by introducing the axial gauge fields to Yang-Mills gauge theories. It is shown that despite of the standard BRST algebra generated by gauge, ghost and anti-ghost fields, this enlarged algebra is closed under BRST and anti-BRST transformations [17]. The necessity of Nakanishi-Lautrup fields [48, 49] is removed here and these auxiliary fields are replaced by a set of functionals of ghosts and anti-ghosts. This functional can be considered as a colored scalar field with null ghost-number. On the other hand, the extra degrees of freedom carried by auxiliary fields are compensated by axial gauge fields in the extended Lagrangian. In fact, this development gives a conceptual framework and a geometric formalism to algebraic anti-BRST transformations and anti-ghosts. This also leads to a geometric description of extended BRST quantization of gauge theories [50, 51].
 \par Using these extended BRST transformations via the approach of [18-20], an extended sequence of descent equations and consequently a modified version of consistent anomaly and of consistent Schwinger term is found. According to [4], none of vector and axial currents is the only responsible for axial anomaly. Particularly, the anomalous behavior of the theory should be studied by considering the contribution of vector and axial currents simultaneously. Thus, it seems more natural to consider modified consistent anomalies and modified consistent Schwinger terms as anomalous behaviors of gauge theories.
 \par In section 1, \emph{Axial Extension of a Gauge Theory}, the idea of axially extended gauge theories is studied in the setting of differential geometry. In second 2, \emph{Extended BRST Transformations}, a summarized review over geometric ghosts and BRST transformations is given and then the geometrical aspects of anti-ghosts and anti-BRST transformations are worked out. Finally, in section 3, \emph{Extended Descent Equations}, the modified version of anomalous terms is derived and studied properly.
\par In fact, in this article it is shown that the following diagram commutes for gauge theories;\\
 \begin{eqnarray}
 \begin{array}{ccc}
   \textbf{\emph{\emph{Classical Gauge Symmetry}}} & \begin{array}{c}
                                \emph{\emph{Quantization}} \\
                                \longrightarrow \\

                              \end{array}
    & \textbf{\emph{\emph{Quantum BRST Symmetry}}} \\
   \emph{\emph{Axial Extension}} \downarrow &   & \downarrow \emph{\emph{Axial Extension}} \\
   \textbf{\emph{\emph{Classical Local Axial Symmetry}}} & \begin{array}{c}
                                  \\
                                \longrightarrow \\
                                 \emph{\emph{Quantization}}
                              \end{array} & \textbf{\emph{\emph{Quantum Anti-BRST Symmetry }}}
 \end{array}
 \end{eqnarray}

 \par Although it was thought that quantization causes the classical gauge symmetry to produce two different quantum symmetries [17], due to BRST and anti-BRST transformations, in this article it is shown the anti-BRST invariance is the quantized version of classical local axial symmetry, which is broken in standard Yang-Mills theories. This broken symmetry revives through quantization and produces the anti-BRST invariance.\\

%%%%%%%%%%%%%%%%%%%%%%%%%%%%%%%%%%%%%%%%

\par
\section{Axial Extension of a Gauge Theory}
\setcounter{equation}{0}
\par Extending a renormalizable field theory is the process of adding a number of new renormalizable terms to the Lagrangian density in order to enlarge its symmetry group. This process may or may not add a number of new fields to the theory. It can be seen that this can replace a global symmetry by a local one. From the perturbation theory point of view this may result in appearance of a number of new Feynman diagrams which will affect the renormalization process and consequently the anomalous behaviors of the theory. In this section, we study the axial extension of gauge theories in a geometric framework. By axially extension we mean enlarging a Yang-Mills gauge theory to produce an axial gauge theory which admits local axial symmetry.\\

%%%%%%%%%%%%%%%%%%%%%%%%%%

\par
\subsection{Axially Extended Gauge Theory}
\setcounter{equation} {0}
\par The main goal of axial extension of a gauge theory is to add a number of new renormalizable terms to the Lagrangian density in order to make the axial symmetry appear in the local form. Thus, generally consider a Yang-Mills gauge theory over  $\mathbb{R}^4$ by;
\begin{eqnarray} \label {1-1-1}
\mathcal{L}=\mathcal{L}_{Yang-Mills}+\mathcal{L}_{Dirac}+A^a_\mu \overline{\psi}\gamma^\mu T^a \psi~,
\end{eqnarray}
\par \noindent with gauge group $G$ and the gauge transformation group $\mathcal{G}$. Under the action of an element $ e^{-i\alpha(x)}\in \mathcal{G}$ with;
\begin{eqnarray} \label {1-1-2}
\begin{array}{c}
  \psi'(x)=e^{-i\alpha(x)}\psi(x) \\
  A'_\mu (x)= e^{-i\alpha(x)} i\partial_\mu e^{i\alpha(x)}+\emph{\emph{Ad}}_e^{-i\alpha(x)}(A_\mu (x))
\end{array}
\end{eqnarray}
\par \noindent $\mathcal{L}$ remains invariant. But the action of axial transformation $e^{-i\alpha(x)\gamma_5}$ on $\mathcal{L}$  yields a nontrivial variation; $\delta \mathcal{L}=e^{-i\alpha(x)\gamma_5} i\partial_\mu e^{i\alpha(x)\gamma_5}$, which cannot be compensated by variations of gauge fields $A_\mu^a (x)$. Essentially, this is the reason that axial transformations are global symmetries; there is no geometric structure for such transformations. To overcome this difficulty one should introduce a number of new axial gauge fields to capture the variations of $\mathcal{L}$ under axial transformations. To this end, a set of new terms, $\overline{\psi}\gamma^\mu \gamma_5 T^a \psi$, each of which coupled to an axial gauge field, say $B_\mu^a$, should be added to $\mathcal{L}$. The resulting theory takes the form of;
\begin{eqnarray} \label {1-1-3}
\mathcal{L}_{ex}=\mathcal{L}_{Gauge}+\mathcal{L}_{Dirac}+A^a_\mu \overline{\psi}\gamma^\mu T^a \psi+B^a_\mu \overline{\psi}\gamma^\mu \gamma_5 T^a \psi~,
\end{eqnarray}
\par \noindent where  $\mathcal{L}_{Gauge}$ is a functional of $A_\mu^a (x)$ and $B_\mu^a (x)$ which will be determined in the following. It can be easily checked that the invariance of $\mathcal{L}_{ex}-\mathcal{L}_{Gauge}$ under gauge transformation $e^{-i\alpha(x)}$ requires (\ref {1-1-2}) together with;
\begin{eqnarray} \label {1-1-4}
B'_\mu (x)=\emph{\emph{Ad}}_{e^{-i\alpha(x)}} (B_\mu (x))~,
\end{eqnarray}
\par \noindent where $B_\mu(x)=B_\mu^a (x) \gamma_5 T^a$. On the other hand, the invariance of $\mathcal{L}_{ex}-\mathcal{L}_{Gauge}$ under axial transformation $e^{-i\alpha(x)\gamma_5}$ requires;
\begin{eqnarray} \label {1-1-5}
\begin{array}{c}
\psi'(x)=e^{-i\alpha(x) \gamma_5} \psi(x)~, \\
  A'_\mu (x)+B'_\mu (x)=e^{-i\alpha(x)\gamma_5} i\partial_\mu e^{i\alpha(x)\gamma_5}+\emph{\emph{Ad}}_{e^{-i\alpha(x)\gamma_5}} (A_\mu(x)+B_\mu(x))~.
\end{array}
\end{eqnarray}
\par Notice that the axial transformation mixes the vector and axial gauge fields. This is a crucial fact which is discussed in the following. To find a gauge/axial invariant Lagrangian density $\mathcal{L}_{ex}$, one should look for a gauge/axial invariant $\mathcal{L}_{Gauge}$. The second part of (\ref {1-1-5}) forces $\mathcal{L}_{Gauge}$ to be a functional of a mixed form of  $A_\mu(x)$ and $B_\mu(x)$. Moreover, according to (\ref {1-1-2}), (\ref {1-1-4}) and (\ref {1-1-5}), the gauge and axial transformations are respectively given by;
\begin{eqnarray} \label {1-1-6}
\begin{array}{c}
  (A_\mu+B_\mu)'(x)=e^{-i\alpha(x)} i\partial_\mu e^{i\alpha(x)} +\emph{\emph{Ad}}_{e^{-i\alpha(x)}}(A_\mu (x)+B_\mu (x))~, \\
  (A_\mu+B_\mu)'(x)=e^{-i\alpha(x)\gamma_5} i\partial_\mu e^{i\alpha(x)\gamma_5} +\emph{\emph{Ad}}_{e^{-i\alpha(x)\gamma_5}}(A_\mu (x)+B_\mu (x))~.
\end{array}
\end{eqnarray}
\par \noindent Both of these transformations are completely similar to the gauge field part of (\ref {1-1-2}) which keeps the pure Yang-Mills Lagrangian density invariant. Thus, if one defines;
\begin{eqnarray} \label {1-1-7}
F_{\mu\nu}=\partial_\mu (A_\nu+B_\nu)-\partial_\nu (A_\mu+B_\mu)-i[(A_\mu+B_\mu),(A_\nu+B_\nu)]~,
\end{eqnarray}
\par \noindent the Yang-Mills Lagrangian density $\mathcal{L}_{Yang-Mills}=-\frac{1}{4} Tr\{F_{\mu\nu} F^{\mu\nu}\}$ would be a compatible candidate for $\mathcal{L}_{Gauge}$. Here we use the normalized trace with $tr~1=1$ for $1=\gamma_5^2$, the $4\times 4$ identity matrix acting on spinors. Note that although the trace of $\gamma_5$ vanishes the Lagrangian density cannot be split into two different parts, each of which a functional of $A_\mu$ or $B_\mu$. We emphasize that in the pure Yang-Mills Lagrangian density, $\mathcal{L}_{Yang-Mills}$, the vector and the axial gauge fields, $A_\mu$ and $B_\mu$, are both considered conventionally in the fundamental representation of the gauge group, independent to their representations on spinors which we show simply with $T^a$ matrices. \\
\par Take the infinitesimal gauge transformation of $e^{-ti\alpha}\in \mathcal{G}$, $t\in \mathbb{R}$;
\begin{eqnarray} \label {1-1-8}
\begin{array}{c}
  \Delta_\alpha A_\mu:= \frac{\emph{\emph{d}}}{\emph{\emph{d}}t}|_{t=0} e^{-it\alpha(x)} i\partial_\mu e^{it\alpha(x)} +\emph{\emph{Ad}}_{e^{-it\alpha(x)}} (A_\mu (x)))=-\partial_\mu \alpha+i[A_\mu,\alpha]~, \\
  \Delta_\alpha B_\mu:=\frac{\emph{\emph{d}}}{\emph{\emph{d}}t}|_{t=0} \emph{\emph{Ad}}_{e^{-it\alpha(x)}} (B_\mu(x))=i[B_\mu,\alpha]~,
\end{array}
\end{eqnarray}
\par \noindent and the infinitesimal axial transformation of $e^{-ti\alpha\gamma_5}$, $t\in \mathbb{R}$;
\begin{eqnarray} \label {1-1-9}
\Delta'_\alpha (A_\mu+B_\mu ):= \frac{\emph{\emph{d}}}{\emph{\emph{d}}t}|_{t=0} e^{-it\alpha(x)\gamma_5} i\partial_\mu e^{it\alpha(x)\gamma_5} +\emph{\emph{Ad}}_{e^{-it\alpha(x)\gamma_5}} (A_\mu (x)+B_\mu (x))~.
\end{eqnarray}
 \par Indeed, it can be seen that the infinitesimal axial transformations of gauge and axial gauge fields are given by;
 \begin{eqnarray} \label {1-1-10}
 \begin{array}{c}
   \Delta'_\alpha A_\mu=i[B_\mu,\alpha\gamma_5]~, \\
   \Delta'_\alpha B_\mu=-\partial_\mu \alpha\gamma_5+i[A_\mu,\alpha\gamma_5]~.
 \end{array}
\end{eqnarray}
 \par \noindent If one defines the vector and axial gauge fields as operator valued 1-forms, then the gauge and axial transformations of $A$ and $B$ are respectively given by the following forms;
\begin{eqnarray} \label {1-1-11}
\begin{array}{c}
  \Delta_\alpha A=-\emph{\emph{d}}\alpha+i[A,\alpha]~, \\
  \Delta_\alpha B=i[B,\alpha]~,
\end{array}
\end{eqnarray}
 \par \noindent and;
\begin{eqnarray} \label {1-1-12}
\begin{array}{c}
  \Delta'_\alpha A=i[B,\alpha\gamma_5]~, \\
  \Delta'_\alpha B=-\emph{\emph{d}}\alpha\gamma_5+i[A,\alpha\gamma_5]~.
\end{array}
\end{eqnarray}
 \par Therefore, an extended Lagrangian density, $\mathcal{L}_{ex}$, is found properly. Moreover, as we expected before, $\mathcal{L}_{ex}$ not only contains $\mathcal{L}$ (the original theory) but it admits the axial symmetry in the local form. Indeed, up to anomalous behaviors, $\mathcal{L}_{ex}$ is a renormalizable theory by power counting. Note that from the perturbation theory point of view the fact of which $\mathcal{L}_{ex}$ contains $\mathcal{L}$, results in the appearance of all Feynman diagrams of the original theory in the perturbative calculations of the extended theory. Thus $\mathcal{L}$ may be considered as a sub-theory for $\mathcal{L}_{ex}$. In the following, the next subsection, it is shown that the extended Lagrangian, $\mathcal{L}_{ex}$, can be considered as a gauge theory with an enlarged gauge group.\\

%%%%%%%%%%%%%%%%%%%%%%%%%%%%%%%%%%%%%%%%%%

\par
\subsection{Extended Gauge Transformation Group}

\par
 As it was stated above, an axially extended gauge theory admits the local axial symmetry, but it has not shown yet that geometrically the extended Lagrangian density is also a gauge theory by itself. To show this fact the geometric theory of gauge fields is revisited now.
 Let $M$ be an even dimensional spin manifold and $G$ a semi-simple (or compact) Lie group with Lie algebra $\mathfrak{g}$. Consider a principal $G$-bundle over $M$;
\begin{eqnarray} \label {1-2-1}
G\hookrightarrow \mathcal{P}\twoheadrightarrow M~,
\end{eqnarray}
\par \noindent and set a connection over $\mathcal{P}$ with the Cartan connection form $\pi$ [52]. Now suppose that $V$ is a complex irreducible representing space of $G$. Also consider a Hermitian inner product for $V$, say $\langle,\rangle$, and assume that the action of $G$ is unitary. The representation of $G$ on $V$ defines a complex vector bundle over $M$ with;
\begin{eqnarray} \label {1-2-2}
V\hookrightarrow E:=\mathcal{P}\times_G V\twoheadrightarrow M~,
\end{eqnarray}
\par \noindent where;
\begin{eqnarray} \label {1-2-3}
\mathcal{P}\times_G V:=\{[(p,v)]|(p,v)\in \mathcal{P}\times V~\emph{\emph{and}}~(p,v)\sim(p\lhd g,g^{-1}\rhd v)\}~.
\end{eqnarray}
\par \noindent It can easily be checked that $E\twoheadrightarrow M$ is a vector bundle and any one of its sections can be represented by an equivalence class, say $[(s,\xi)]$, for some $s\in \Gamma(P)$ and $\xi\in C^\infty (M,V)$. Clearly, by definition (\ref {1-2-3}), the choices of $s$ and $\xi$ are not unique. As it will be discussed in the following, these different choices produce the gauge symmetry. On the other hand, the connection of $\mathcal{P}$ induces a connection over $E$ with covariant derivative;
\begin{eqnarray} \label {1-2-4}
\nabla_X [(s,\xi)]:=[(s,\emph{\emph{d}}\xi(X))]+[(s,\pi(\emph{\emph{d}}s(X))\rhd \xi)]~,
\end{eqnarray}
\par \noindent where $\pi(ds(X))\rhd \xi$ is the action of $\mathfrak{g}$, induced by the representation of $G$ (for any $T\in \mathfrak{g}$ and $v\in V$, define; $T\rhd v:= \frac{\emph{\emph{d}}}{\emph{\emph{d}}t}|_{t=0} e^{tT}\rhd v)$. The definition (\ref {1-2-4}) is independent of the choices $s$ and $\xi$. Moreover, the pull back of $\pi$ through $s$, $s^* (\pi)$, defines an operator valued 1-form over $M$. Conventionally, $s^* (\pi)$ is called the connection form. Basically these operator valued 1-forms play the roles of gauge fields in the context of Yang-Mills gauge theories. To see this note that;
\begin{eqnarray} \label {1-2-5}
 (s\rhd g)^*(\pi)=g^{-1}\emph{\emph{d}}g+\emph{\emph{Ad}}_{g^{-1}}(s^* (\pi))~,
\end{eqnarray}
\par \noindent which is thoroughly similar to the gauge field part of (\ref {1-1-2}) with replacing $s^* (\pi)$ (resp. $(s\rhd g)^*(\pi)$) with; $-iA=-iA_\mu dx^\mu$ (resp. $-iA'=-iA'_\mu dx^\mu$).\\
 \par Now consider the tensor product bundle;
\begin{eqnarray} \label {1-2-6}
\mathcal{S}\otimes V\hookrightarrow \mathcal{S}(M)\otimes E\twoheadrightarrow M
\end{eqnarray}
\par \noindent for the spin bundle $\mathcal{S}(M)\twoheadrightarrow M$ and its standard fiber $\mathcal{S}$. The definition (\ref {1-2-4}) and the connection of the spin bundle naturally induce a covariant derivative, $\nabla_\otimes$, over the tensor product bundle (\ref {1-2-6}). Thus if $\mathcal{D}$ is the Dirac operator of $\nabla_\otimes$, then the Lagrangian density of matter is given by;
\begin{eqnarray} \label {1-2-7}
\mathcal{L}_{Matter} (\psi,\mathcal{D})=\langle\overline{\psi},\mathcal{D}\psi\rangle
\end{eqnarray}
\par \noindent for $\Gamma(\mathcal{S}(M)\otimes E)$. According to definitions (\ref {1-2-3}) and (\ref {1-2-4}), it can be seen that $\mathcal{L}_{Matter} (\psi,D)$ is a gauge invariant functional. The other gauge invariant functional on 4-dimensional $M$ with Minkowsian signature is the Yang-Mills lagrangian density;
\begin{eqnarray} \label {1-2-8}
\mathcal{L}_{Yang-Mills}(\nabla_\otimes ):=\frac{1}{2}Tr\{R\wedge\bigstar R\}~,
\end{eqnarray}
\par \noindent with $R$ the curvature tensor of  $\nabla_\otimes$ and with $\bigstar$ the Hodge star operator. When $M=\mathbb{R}^4$, the spin bundle is trivial and thus it is seen from (\ref {1-2-4}) that; $R=\emph{\emph{d}}s^* (\pi)+s^*(\pi)^2$, where here, as mentioned above, $R$ is written in the fundamental representation. Usually, $\mathcal{L}_{Yang-Mills}$ is defined by referring to $F=2iR$. Thus, one may conventionally define; $\mathcal{L}_{Yang-Mills} (\nabla_\otimes)=-\frac{1}{8} Tr\{F\wedge\bigstar F\}$ \footnote{Here we recall that the Hodge dual of a differential $m$-form over a pseudo-Riemannian $(m+n)$-dimensional manifold, locally is defined with;
$\bigstar\omega:=\frac{\sqrt{|\det g|}}{n!}\epsilon_{\mu_1,...,\mu_m,\nu_1,...,\nu_n}\omega^{\mu_1,...,\mu_m}\emph{\emph{d}}x^{\nu_1}\wedge...\wedge \emph{\emph{d}}x^{\nu_n}$, for $\omega$ a differential $m$-form locally given by
$\omega_{\mu_1,...,\mu_m}\emph{\emph{d}}x^{\mu_1}\wedge...\wedge \emph{\emph{d}}x^{\mu_m}$, over local coordinates $\{x^\mu\}_{\mu=1}^{m+n}$, for totally anti-symmetric components $\omega_{\mu_1,...,\mu_m}$, $1\leq\mu_1,...,\mu_m\leq m+n$. Therefore, $-\frac{1}{8} Tr\{F\wedge\bigstar F\}$ is essentially equal to $-\frac{1}{4}Tr\{F_{\mu\nu}F^{\mu\nu}\}$}.\\
\par Now the axially extended gauge theories can be studied by these geometric structures. First, note that for any given $d$-dimensional Lie algebra $\mathfrak{g}$ there exists a natural way to define a $2d$-dimensional Lie algebra $\tilde{\mathfrak{g}}$ which contains $\mathfrak{g}$ as a Lie sub-algebra. To see this, chose a basis for $\mathfrak{g}$, say $\{t^a\}_{a=1}^d$, and define a set of new elements $\{s^a\}_{a=1}^d$ together with the following commutation relations;
\begin{eqnarray} \label {1-2-9}
  [s^a,s^b]=[t^a,t^b]=C^{abc} t^c ~~~~~~,~~~~~~~[s^a,t^b]=-[t^b,s^a]=C^{abc} s^c~.
\end{eqnarray}
\par \noindent It can be checked easily that the Jacobi identity is satisfied by the relations of (\ref {1-2-9}) and thus they form a Lie algebra, $\tilde{\mathfrak{g}}$. Indeed $\tilde{\mathfrak{g}}$ can be considered as an extension of $\mathfrak{g}$. However, it is clear that this extension is well-defined, unique, and natural; i.e. if $\tilde{\mathfrak{g}}'$ is constructed over $\mathfrak{g}$ for another basis, say $\{t'^a\}_{a=1}^d$, then $\tilde{\mathfrak{g}}'\cong \tilde{\mathfrak{g}}$.
\par \noindent It is clear that for a given representation of $\mathfrak{g}$, say $\phi$, and for a nontrivial involutive matrix $\gamma$ which commutes with all $\phi(t^a)$s, one can produce a representation of $\tilde{\mathfrak{g}}$, say $\tilde{\phi}$. It is enough to set $\tilde{\phi}(s^a)=\gamma\phi(t^a)$. This can be done with $\gamma=\gamma_5$ for any representation of the Lie algebra $\mathfrak{g}$ on the spinors over an even dimensional spin manifold. Thus, we refer to the extension procedure of (\ref {1-2-9}) as axial extension of a Lie algebra.
 \par As it was stated above, $\tilde{\mathfrak{g}}$ is itself a Lie algebra and thus there exists a simply connected Lie group $\tilde{G}$ with $\emph{\emph{Lie}}\tilde{G}=\tilde{\mathfrak{g}}$. Since $\tilde{\mathfrak{g}}$ contains $\mathfrak{g}$ as a Lie sub-algebra, then there is a closed Lie subgroup of $\tilde{G}$, with Lie algebra $\mathfrak{g}$. Actually, $\tilde{G}$ contains a simply connected Lie subgroup, say $G'$, with $\emph{\emph{Lie}} G'=\mathfrak{g}$. Indeed, $\tilde{G}\cong G'\times G'$. To see this note that $\{t_R^a:=\frac{1}{2} (t^a+s^a),t_L^a:=\frac{1}{2}(t^a-s^a)\}_{a=1}^d$, forms a basis for $\tilde{\mathfrak{g}}$  with the following Lie brackets;
\begin{eqnarray} \label {1-2-10}
  \begin{array}{c}
    [t_R^a,t_R^b]=C^{abc} t_R^c~,~~
    [t_L^a,t_L^b]=C^{abc} t_L^c~,~~
    [t_L^a,t_R^b]=0~.
  \end{array}
\end{eqnarray}
\par \noindent This shows that $\tilde{\mathfrak{g}}$ can be decomposed into the direct sum of two copies of $\mathfrak{g}$, $\tilde{\mathfrak{g}}=\mathfrak{g}\oplus\mathfrak{g}$. On the other hand, since $\pi_1 (G'\times G')=\pi_1 (G')\times \pi_1 (G')$, it follows that $\tilde{G}=G'\times G'$, and thus the axial extension of any Lie algebra induces an extension of its simply connected Lie group. Therefore, the axial extension of $su(N)$ will induce an extension of $SU(N)$. Hence, the axial extension of a $SU(N)$-gauge theory leads to a $SU(N)_L \times SU(N)_R$-gauge theory.
\par Generally, for simply connected $G$ and $\tilde{G}$ the inclusion of $i:\mathfrak{g}\hookrightarrow\tilde{\mathfrak{g}}$ defines a natural inclusion of $G$ into $\tilde{G}$ denoted by $I:G\hookrightarrow\tilde{G}$, with $\emph{\emph{d}}I=i$ [53]. Thus, if $\mathcal{P}\twoheadrightarrow M$ and $\tilde{\mathcal{P}}\twoheadrightarrow M$ are principal bundles over $M$ for simply connected Lie groups $G$ and $\tilde{G}$ respectively, and if $\xi:\mathcal{P}\rightarrow \tilde{\mathcal{P}}$ is a principal bundle homomorphism which by definition makes the following diagram commutative
\begin{eqnarray} \label {1-2-11}
\begin{array}{ccccc}
  \mathcal{P}\times G & \rightarrow & ~~~\mathcal{P} & \twoheadrightarrow & ~~~~M \\
  \xi~\downarrow ~~~\downarrow~I &   & \xi~\downarrow & ~~~ & =~\downarrow \\
  \tilde{\mathcal{P}}\times \tilde{G} & \rightarrow & ~~~\tilde{\mathcal{P}} & \twoheadrightarrow & ~~~~M
\end{array}
~,
\end{eqnarray}
\par \noindent then a natural reduction of $\tilde{G}$-gauge theories to $G$-gauge theories over $M$ is defined through $\xi$. It is enough to set the Cartan connection form over $\mathcal{P}\twoheadrightarrow M$, the pull back of that over $\tilde{\mathcal{P}}\twoheadrightarrow M$, say $\tilde{\pi}$, through $\xi$, i.e. $\pi:=\xi^* (\tilde{\pi})$. Such a reduction procedure is always possible for simply connected $M$, e.g. $M=\mathbb{R}^4$. In this sense, the reduction procedure can be considered as the inverse project of extension gauge theories.
\par As it was stated above, one can consider $\{T^a,T^a \gamma_5\}_{a=1}^d$ as a representing basis of the extended Lie algebra $\mathfrak{g}$ over the spinors. With this notation, (\ref {1-2-9}) also confirms the infinitesimal gauge and axial transformations of (\ref {1-1-11}) and (\ref {1-1-12}). More precisely, (\ref {1-1-3}) defines an ordinary $\tilde{G}$-gauge theory. Consequently, the gauge and axial transformations for $\mathcal{L}$ are essentially the gauge transformations for $\mathcal{L}_{ex}$, the extended Lagrangian density. These transformations are conventionally referred to as extended gauge transformations.
\par The relations of (\ref {1-2-10}) assert that one can consider a mixed form of $A$ and $B$ to define a number of new gauge fields which do not mix under extended gauge transformations. To this end set;
\begin{eqnarray} \label {1-2-12}
A_R=(\frac{A+B}{2}) P_R~~~,~~~A_L=(\frac{A+B}{2}) P_L~,
\end{eqnarray}
\par \noindent where $P_R=\frac{1+\gamma_5}{2}$ and $P_L=\frac{1-\gamma_5}{2}$ are respectively the projections onto right and left handed Weyl spinors due to chiral fermions. Thus the infinitesimal transformations take the following form;
\begin{eqnarray} \label {1-2-13}\begin{array}{c}
                                  \Delta^R_\alpha A_R=-\frac{1}{2} \emph{\emph{d}}\alpha_R+i[A_R,\alpha_R]~,~~ \\
                                  \Delta^L_\alpha A_L=-\frac{1}{2} \emph{\emph{d}}\alpha_L+i[A_L,\alpha_L]~,~~~ \\
                                  \Delta^R_\alpha A_L=\Delta^L_\alpha A_R=0~,~~~~~~~~~~~~~
                                \end{array}
\end{eqnarray}
\par \noindent for $\Delta^R$ (resp. $\Delta^L$), the right- (resp. left-) handed infinitesimal chiral transformation. This splits the gauge fields into right- and left-handed components.\\

%%%%%%%%%%%%%%%%%%%%%%%%%%%%%%%%%%%%%%%%%%%%
\par
\section{Extended BRST Transformations}
\setcounter{equation} {0}
\par  It is well-known that the BRST invariance is the quantized version of the classical gauge symmetry in the context of quantum Yang-Mills gauge theories [12-14]. Actually, the BRST invariance can reproduce all the information of gauge symmetries within the quantized formalism. In this section the geometric structure of anti-BRST transformations and consequently of anti-ghost is given by using the geometric ideas of BRST quantization. In the first subsection, a review over the geometrical theory of ghosts and BRST transformations is given.\\

%%%%%%%%%%%%%%%%%%%%%%%%%%%%%%%%%%%%%%%%%%%%%

\par
\subsection{Ghosts and BRST transformations, a Geometric Approach}
\setcounter{equation} {0}
\par The geometric description of Fddeev-Popov quantization, not only illustrates the canonical structures of ghost fields and BRST transformations, but it results in simple proofs for their substantial behaviors [37-39]. To see this, assume that $M$, the space-time, is an even dimensional spin manifold. Consider the principal $G$-bundle of (\ref {1-2-1}) and the Cartan connection form $\pi$ over it. The space of all connection forms is an Affine space, denoted by $\mathcal{A}$. The gauge transformation defines a right action of the infinite dimensional Lie group $C^\infty (M,G)$ on $\mathcal{A}$. For example the action of $g\in C^\infty (M,G)$ on $A\in \mathcal{A}$ is given by;
\begin{eqnarray} \label {2-1-1}
A\lhd g=ig^{-1} \emph{\emph{d}}g+\emph{\emph{Ad}}_{g^{-1}} (A)~.
\end{eqnarray}
 \par Consider $\mathcal{G}\in C^\infty (M,G)$ to be the set of base point preserving elements [37]. Then, the free action of $\mathcal{G}$ on  $\mathcal{A}$ defines a  principal $\mathcal{G}$-bundle;
\begin{eqnarray} \label {2-1-2}
\mathcal{G}\hookrightarrow \mathcal{A}\twoheadrightarrow \mathcal{A}/\mathcal{G}~.
\end{eqnarray}
 \par \noindent Fix a connection over this principal bundle and denote its Cartan connection form by $\Pi$. For a fixed $A_0\in \mathcal{A}$ define the fiber map;
\begin{eqnarray} \label {2-1-3}
\begin{array}{c}
  i_{A_0}:M\times \mathcal{G}\rightarrow M\times \mathcal{A}~,~~~~ \\
  i_{A_0}:(m,g)\mapsto(m,A_0\lhd g)~.
\end{array}
\end{eqnarray}
\par Define $\mathfrak{A}$ as a $\mathfrak{g}$-valued 1-form over $M\times \mathcal{A}$ with;
\begin{eqnarray} \label {2-1-4}
\mathfrak{A}(v_m,\eta_A )=A(v_m)-i\Pi(\eta_A )(m)~,
\end{eqnarray}
\par \noindent for $(v_m,\eta_A )\in T_m M\times T_A \mathcal{A}$. Then set $A+\omega:=i_{A_0}^* (\mathfrak{A})$, where $A$ and $\omega$ are the pull backs of the first and the second parts of $\mathfrak{A}$ respectively. Conventionally $\omega$ is called ghost. Indeed, $\omega$ is a $C^\infty (M,\mathfrak{g})$-valued left invariant 1-form over $\mathcal{G}$ and then its color components anti-commute with each other as simple 1-forms. Thus, its color components behave like Grassmannian numbers in path-integral frameworks as group indexed ghosts do.\\
 \par Denote the exterior derivative operator over $\mathcal{A}$ by $\emph{\emph{d}}_{\mathcal{A}}$ and define $\delta$ to be its pull back through $i_{A}^*$. Then a direct calculation gives;
\begin{eqnarray} \label {2-1-5}
\begin{array}{c}
 \delta A=\emph{\emph{d}}\omega-iA\omega-i\omega A~, \\
 \delta \omega=-i\omega^2 ~.~~~~~~~~~~~~~~~
\end{array}
\end{eqnarray}
 \par \noindent Moreover, since $(\emph{\emph{d}}+\emph{\emph{d}}_{\mathcal{A}})^2=\emph{\emph{d}}_{\mathcal{A}}^2=0$, we find that; $\delta^2=0$ and $\delta \emph{\emph{d}}+\emph{\emph{d}} \delta=0$, similar to ordinary BRST derivation. This is the reason for $\omega$ and $\delta$ to be respectively considered as the standard ghost and BRST derivative [37]. In fact, for better understanding (\ref {2-1-5}), it should be compared with the vector gauge field part of (\ref {1-1-11}). Indeed, any 1-parameter group gauge transformation, say $e^{it\alpha}$, $t\in \mathbb{R}$, induces a fiber-wise vector field over $\mathcal{A}$, shown by $\eta^\alpha$, given by
\begin{eqnarray} \label {2-1-6}
\eta_A^\alpha=\sum_{a,\mu=1}^{d,D}\int_{x\in M}(\Delta_\alpha A)_\mu^a(x) \frac{\delta}{\delta A_\mu^a(x)}\in T_A \mathcal{A}~ ,
\end{eqnarray}
\par \noindent for $\{A_\mu^a (x)\}_{a,\mu=1,x\in M}^{d,D}$ ($d=\dim \mathfrak{g}$  and  $D=\dim M$) the coordinate functions of $\mathcal{A}$. Moreover, $\delta A$ in (\ref {2-1-5}), is a differential 2-form on $M\times \mathcal{G}$ with;
\begin{eqnarray} \label {2-1-7}
i_{\eta^\alpha} \delta A=i_{\eta^\alpha} (\emph{\emph{d}}\omega-iA\omega-i\omega A)=-\emph{\emph{d}}\alpha+i[A,\alpha]=\Delta_\alpha A~,
\end{eqnarray}
\par \noindent for $i_{\eta^\alpha}$ the internal multiplier operator [53]. More generally, infinitesimal gauge transformations are infinitesimal moves through the fibers of $\mathcal{A}$ [37] and therefore, they are smooth sections of tangent bundle $T\mathcal{A}$. Basically, infinitesimal gauge transformations are fiber-wise vector fields on $\mathcal{A}$. On the other hand, ghosts are fiber-wise 1-forms over $\mathcal{A}$ and thus, they can evaluate the infinitesimal gauge transformations as their dual objects. In fact, the equivalence of gauge symmetry and BRST invariance can be illustrated by noting that the former is defined with the elements of $T\mathcal{A}$ but the later one is given in terms of their dual objects in the cotangent bundle, $T^* \mathcal{A}$. It is the same idea of Legendre transformation in classical mechanics which translates the Lagrangian formalism into the Hamiltonian formulation over a symplectic manifold. This intuition is one of the cornerstones for geometric approach to BRST quantization.\\

%%%%%%%%%%%%%%%%%%%%%%%%%%%%%%%%%%%%%%%%%%%%%

\par
\subsection{Anti-Ghosts and Anti-BRST Transformations}
\par As stated above, the extension of gauge group enlarges the algebra of group indexed ghosts. In fact, in this subsection it is shown that how the anti-ghost fields emerge through the axial extension of a Yang-Mills gauge theory. Consider $\mathcal{A}$ and $\mathcal{A}_5$ as the set of all vector and axial connection forms respectively, and take the space of extended connection forms as the Cartesian product space $\mathcal{A}\times \mathcal{A}_5$. Then consider $\tilde{\mathfrak{g}}$, $\tilde{G}$ and $\tilde{\mathcal{G}}$ as the axially extended versions of Lie algebra $\mathfrak{g}$, gauge group $G$, and gauge transformation group $\mathcal{G}$, respectively. Finally, define $\mathfrak{A}$ and $\Pi$ similar to (\ref {2-1-4}) and let $P:\tilde{\mathfrak{g}}\twoheadrightarrow \mathfrak{g}$ and $P_5:=1-P$ be two projections. The fiber map for a fixed point $(A_0,B_0)\in \mathcal{A}\times \mathcal{A}_5$ is given by;
\begin{eqnarray} \label {2-2-1}
\begin{array}{c}
  i_{(A_0,B_0)}:M\times \tilde {\mathcal{G}} \rightarrow M\times \mathcal{A}\times \mathcal{A}_5~,~~~ \\
  i_{(A_0,B_0)}:(m,g)\mapsto(m,(A_0,B_0 )\lhd g)  .
\end{array}
\end{eqnarray}
\par \noindent Now define $A+B+\omega+\omega^*:=i_{(A_0,B_0)}^*(\mathfrak{A})$, with $\omega :=i_{(A_0,B_0)}^* (P(-i\Pi))$ and $\omega^*:=i_{(A_0,B_0)}^* (P_5(-i\Pi))$. Thus if $\omega$ and $\omega^*$ are the pull backs of the exterior derivative operators on $\mathcal{A}$, $\emph{\emph{d}}_A$, and on $\mathcal{A}_5$, $\emph{\emph{d}}_{\mathcal{A}_5}$, respectively, then it can similarly be shown that\footnote{One can simply apply the notation of super-commutator for elements of non-trivial degrees such as (operator valued) differential forms; i.e. $[a,b]:=ab-(-1)^{|a||b|}ba$. Therefore, for example we write; $\delta A=\emph{\emph{d}}\omega-i[A,\omega]$. Here and in the following, to have more clear formulas we refuse this super-algebraic notation.};
\begin{eqnarray} \label {2-2-2}
\begin{array}{cc}
  \delta A=\emph{\emph{d}}\omega-iA\omega-i\omega A~, & \delta \omega=-i\omega^2~,~~~~~~~ \\
  \delta B=-iB\omega-i\omega B~,~~~~ & ~~~~~~\delta \omega^*=-i\omega^*\omega-i\omega\omega^*~,
\end{array}
\end{eqnarray}
\par \noindent which is in complete agreement with BRST derivative, replacing the Nakanishi-Lautrup field $h$ [48, 49, 54] by $-i\omega^*\omega-i\omega\omega^*$. Note that $h$ is a colored scalar field with ghost number zero similar to $-i\omega^*\omega-i\omega\omega^*$. On the other hand, as it will be shown in the following $\delta^2$ vanishes and this implies that $\delta$ annihilates both $h$ and $-i\omega^*\omega-i\omega\omega^*$ similarly. Thus, $-i\omega^*\omega-i\omega\omega^*$ can be considered as the Nakanishi-Lautrup field which appears in the path integral quantization formalism. Indeed, it is seen that the ordinary BRST derivation is not closed for classical fields, and group indexed ghosts and anti-ghosts [13, 14, 17], but this new fashion of BRST algebra is closed by itself. Actually, the auxiliary field is replaced by $-i\omega^*\omega-i\omega\omega^*$ in (\ref {2-2-2}) which closes the BRST algebra. Moreover, the degrees of freedom carried by the auxiliary field $h$ are compensated with those due to axial gauge fields. However, $\delta$ is also called the BRST operator while $\omega$ and $\omega^*$ are referred to as ghost and anti-ghost fields, respectively. This gives an elegant geometric description to group indexed anti-ghosts in terms of 1-forms dual to infinitesimal axial transformations.\\
\par \noindent On the other hand, one can similarly show that;
\begin{eqnarray} \label {2-2-3}
\begin{array}{cc}
  \delta^* A=-iB\omega^*-i\omega^* B~,~~~~~ & \delta^* \omega=0~, \\
  \delta^* B=\emph{\emph{d}}\omega^*-iA\omega^*-i\omega^* A~, & ~~~~~~~~\delta^* \omega^*=-i\omega^{*2}~.
\end{array}
\end{eqnarray}
\par \noindent Actually, since $\emph{\emph{d}}_\mathcal{A}^2=\emph{\emph{d}}_{A_5}^2=0$, then $\delta^2=\delta^{*2}=0$. According to (\ref {2-2-3}), $\delta^*$ gives hand a set of transformations which are similar to BRST ones and keep the quantum Lagrangian invariant. Clearly $\delta^*$ is in complete agreement with the algebraic anti-BRST transformation [15-17]. Thus, conventionally $\delta^*$ is called the anti-BRST derivative. Moreover, from $(\emph{\emph{d}}+\emph{\emph{d}}_\mathcal{A})^2=
(\emph{\emph{d}}+\emph{\emph{d}}_{\mathcal{A}_5})^2=(\emph{\emph{d}}_\mathcal{A}+\emph{\emph{d}}_{\mathcal{A}_5})^2=0$, one simply finds;
\begin{eqnarray} \label {2-2-4}
\delta \emph{\emph{d}}+\emph{\emph{d}}\delta=\delta^* \emph{\emph{d}}+\emph{\emph{d}}\delta^*=\delta\delta^*+\delta^*\delta=0~.
\end{eqnarray}
 These anti-commutation relations will result in a generalized form of descent equations [18-20] by using $\emph{\emph{d}}$, $\delta$ and $\delta^*$ alternatively. Note that, $\delta^* \omega$ is given in term of the auxiliary field in the context of the standard anti-BRST transformation [15-17], thus, (\ref {2-2-3}) equalizes automatically the Nakanishi-Lautrup field and $-i\omega^* \omega-i\omega\omega^*$, which confirms the last part of (\ref {2-2-2}). We refer to the union of (\ref {2-2-2}) and (\ref {2-2-3}) as extended BRST transformation or extended BRST derivation. Consequently, it was shown by (\ref {2-2-2})-(\ref {2-2-4}) that the anti-BRST invariance is the quantized counterpart of the local axial symmetry.

%%%%%%%%%%%%%%%%%%%%%%%%%%%%%%%%%%%%%%%%

%%%%%%%%%%%%%%%%%%%%%%%%%%%%%%%%%%%%%%%%%%%%%%%%%%%%%%%%%%%
\section{Extended Descent Equations}
%%%%%%%%%%%%%%%%%%%%%%%%%%%%%%%%%%%%%%%%%%%%%%%%%%%%%%%%%%%
\setcounter{equation}{0}

%%%%%%%%%%%%%%%%%%%%%%%%%%%%%%%%%%%%%%%

 \par\indent  In this section the extended BRST transformation is used to generalize the descent equations [18-20] in a concrete manner. Indeed, this process leads to a modification of consistent anomalous terms in gauge theories. In the other words, the enlargement of gauge group yields a set of generalized form of anomalous terms including consistent anomalies and consistent Schwinger terms via extended BRST derivative. The result will be delivered in a lattice diagram of differential forms which commutes up to deRham exact forms.
%%%%%%%%%%%%%%%%%%%%%%%%%%%%%%%%%%%%%%%%%%

\par
\subsection{Analytic BRST and Anti-BRST Transformations}
\setcounter{equation} {0}
\par The smooth action of a Lie group $G$ on a given smooth manifold $M$ will induce a set of vector fields over $M$. Indeed, if $X$ is a left invariant vector field over $G$ and $e^{tX}$, $t\in \mathbb{R}$, is its integral curve, then for any $m\in M$, $\gamma_m (t):=m\lhd e^{tX}$, is a smooth curve in $M$. Actually, the collection of all vectors $\tilde{X}_m:=\emph{\emph{d}}\gamma_m (\frac{\emph{\emph{d}}}{\emph{\emph{d}}t}|_{t=o})$ defines a smooth vector field over $M$ with their integral curves $\gamma_m$, $m\in M$. Thus, the action of  $\tilde{\mathcal{G}}$ on $\mathcal{A}\times\mathcal{A}_5$ can be considered as a collection of fiber-wise smooth vector fields over $\mathcal{A}\times\mathcal{A}_5$. To formulate this idea analytically, consider a coordinate system over $\mathcal{A}\times\mathcal{A}_5$, say $\{A_\mu^a(x),B_\mu^a(x)\}_{a,\mu=1,x\in M}^{d,D}$, with $d=\dim \mathfrak{g}$ and  $D=\dim M$. Thus, the gauge transformation $e^{it\alpha}$ induces a vector field, $\mathcal{X}^\alpha$, over $\mathcal{A}\times\mathcal{A}_5$ with;
\begin{eqnarray} \label {3-1-1}
\mathcal{X}^\alpha=
\sum_{a,b,c,\mu}\int_M\{(-\partial_\mu \alpha^a (x)-C^{abc} \alpha^b (x)A_\mu^c (x))\frac{\delta}{\delta A_\mu ^a (x)}+(-C^{abc} \alpha^b (x)B_\mu^c (x))\frac{\delta}{\delta B_\mu^a (x)}\}~.
\end{eqnarray}
\par \noindent A generalized form of $\eta^\alpha$ in (\ref {2-1-6}). Moreover, the axial transformation $e^{it\alpha\gamma_5}$ defines another vector field, $\mathcal{Y}^\alpha$, over $\mathcal{A}\times\mathcal{A}_5$ with;
\begin{eqnarray} \label {3-1-2}
\mathcal{Y}^\alpha=
\sum_{a,b,c,\mu}\int_M\{(-C^{abc} \alpha^b (x) B_\mu^c (x))\frac{\delta}{\delta A_\mu ^a (x)}+(-\partial_\mu \alpha^a (x)-C^{abc} \alpha^b (x) A_\mu^c (x))\frac{\delta}{\delta B_\mu^a (x)}\}~,
\end{eqnarray}
\par \noindent where it is supposed that;
\begin{eqnarray} \label {3-1-3}
\begin{array}{c}
  \frac{\delta}{\delta A_\mu^a (x)} A_\nu^b (y)=\delta_b^a~\delta_\nu^\mu~\delta^{(D)}(x-y)~, \\
  \frac{\delta}{\delta B_\mu^a (x)} B_\nu^b (y)=\delta_b^a~\delta_\nu^\mu~\delta^{(D)}(x-y)~, \\
  \frac{\delta}{\delta A_\mu^a (x)} B_\nu^b (y)=\frac{\delta}{\delta B_\mu^a (x)} A_\nu^b (y)=0~,~
\end{array}
\end{eqnarray}
\par \noindent since $\{\frac{\delta}{\delta A_\mu^a (x)},\frac{\delta}{\delta B_\mu^a (x)}\}_{a,\mu=1,x\in M}^{d,D}$ form a basis for tangent spaces of $\mathcal{A}\times\mathcal{A}_5$. Using these notations the infinitesimal gauge transformations of $A_0$ and $B_0$ as vector and axial gauge fields with respect to $e^{it\alpha}$, i.e. (\ref {1-1-8}), are given by;
\begin{eqnarray} \label {3-1-4}
\begin{array}{c}
  (\Delta_\alpha A_0)_\mu^a (x)=A_\mu^a (x)(\Delta_\alpha A_0)=\mathcal{X}^\alpha_{A_0} (A_\mu^a (x))~, \\
  (\Delta_\alpha B_0)_\mu^a (x)=B_\mu^a (x)(\Delta_\alpha B_0)=\mathcal{X}^\alpha_{B_0} (B_\mu^a (x))~.
\end{array}
\end{eqnarray}
\par \noindent provided $\Delta_\alpha A_0$ and $\Delta_\alpha B_0$ be considered as two elements of $\mathcal{A}$ and $\mathcal{A}_5$ respectively. While the infinitesimal axial transformations of $A_0$ and $B_0$ as vector and axial gauge fields with respect to $e^{it\alpha\gamma_5}$, i.e. (\ref {1-1-10}), are given by;
\begin{eqnarray} \label {3-1-5}
\begin{array}{c}
  (\Delta'_\alpha A_0)_\mu^a (x)=A_\mu^a (x)(\Delta'_\alpha A_0)=\mathcal{Y}^\alpha_{A_0} (A_\mu^a (x))~, \\
  (\Delta'_\alpha B_0)_\mu^a (x)=B_\mu^a (x)(\Delta'_\alpha B_0)=\mathcal{Y}^\alpha_{B_0} (B_\mu^a (x))~.
\end{array}
\end{eqnarray}
\par \noindent Here $\Delta'_\alpha A_0$ and $\Delta'_\alpha B_0$ are also considered as two elements of $\mathcal{A}$ and $\mathcal{A}_5$ respectively. In order to extract the analytic forms of BRST and of anti-BRST operators one should go from the tangent bundle $T(\mathcal{A}\times\mathcal{A}_5)$ to its dual $T^*(\mathcal{A}\times\mathcal{A}_5)$ and use the differential form alternatives. In fact, the Cartan connection form $\Pi$, makes it possible to translate the fiber-wise tangential formalisms to the cotangential ones. According to (\ref {3-1-1}) and (\ref {3-1-2}) one finds;
\begin{eqnarray} \label {3-1-6}
\begin{array}{c}
  \mathcal{Z} (\Gamma)=(\emph{\emph{d}}_{\mathcal{A}}\Gamma+\emph{\emph{d}}_{\mathcal{A}_5}\Gamma)(\mathcal{Z})= \\
  \sum\int_M({\delta\Gamma}/{\delta A_\mu^a (x)})
\{i\partial_\mu P(\Pi^a(\mathcal{Z}))(x)-iC^{abc} A_\mu^b(x)P(\Pi^c(\mathcal{Z}))(x)-iC^{abc} B_\mu^b(x)P_5(\Pi^c(\mathcal{Z}))(x)\} \\
+\sum\int_M({\delta\Gamma}/{\delta B_\mu^a (x)})
\{i\partial_\mu P_5(\Pi^a(\mathcal{Z}))(x)-iC^{abc} A_\mu^b (x)P_5(\Pi^c(\mathcal{Z}))(x)-iC^{abc}B_\mu^b(x)P(\Pi^c(\mathcal{Z}))(x)\}~,
\end{array}
\end{eqnarray}
\par \noindent where the sum is over group and space indices, $a,b,c$ and $\mu$, while $\mathcal{Z}\in \emph{\emph{span}}_\mathbb{C} \{\mathcal{X}^\alpha,\mathcal{Y}^\beta\}_{\alpha,\beta}$ acts on $\Gamma\in C^\infty (\mathcal{A}\times\mathcal{A}_5)$. Consequently, taking the pull back of (\ref {3-1-6}) via the fiber map $i_{(A_0,B_0)}$ leads to;
\begin{eqnarray} \label {3-1-7}
\begin{array}{c}
  \mathcal{Z} (\Gamma)=(\delta\Gamma+\delta^*\Gamma)(\mathcal{Z})= \\
  \sum_{a,b,c,\mu}\int_M({\delta\Gamma}/{\delta A_\mu^a (x)})
\{-\partial_\mu \omega^a(\mathcal{Z})(x)+C^{abc} A_\mu^b(x)\omega^c(\mathcal{Z})(x)+C^{abc} B_\mu^b(x)\omega^{*c}(\mathcal{Z})(x)\}~~~ \\
+\sum_{a,b,c,\mu}\int_M({\delta\Gamma}/{\delta B_\mu^a (x)})
\{-\partial_\mu \omega^{*a}(\mathcal{Z})(x)+C^{abc} A_\mu^b (x)\omega^{*c}(\mathcal{Z})(x)+C^{abc}B_\mu^b(x)\omega^c(\mathcal{Z})(x)\}~,
\end{array}
\end{eqnarray}
\par \noindent for $\mathcal{Z}$ any arbitrary left invariant vector field on $\tilde{\mathcal{G}}$. But since each fiber of $\mathcal{A}\times\mathcal{A}_5$ is precisely a copy of $\tilde{\mathcal{G}}$, then any vertical vector of the tangent space $T_{(A_0,B_0)} (\mathcal{A}\times\mathcal{A}_5)$ is an element of $\emph{\emph{span}}_\mathbb{C} \{\mathcal{X}^\alpha|_{(A_0,B_0)},\mathcal{Y}^\beta|_{(A_0,B_0)}\}_{\alpha,\beta}$. Thus, it can be seen that,
\begin{eqnarray} \label {3-1-8}
\begin{array}{c}
\sum_{a,b,c,\mu}\int_M
\{-\partial_\mu \omega^a(x)+C^{abc} A_\mu^b(x)\omega^c(x)+C^{abc} B_\mu^b(x)\omega^{*c}(x)\}({\delta}/{\delta A_\mu^a (x)})~~~ \\
+\sum_{a,b,c,\mu}\int_M
\{-\partial_\mu \omega^{*a}(x)+C^{abc} A_\mu^b (x)\omega^{*c}(x)+C^{abc}B_\mu^b(x)\omega^c(x)\}({\delta}/{\delta B_\mu^a (x)})~,
\end{array}
\end{eqnarray}
\par \noindent is the fiber-wise exterior derivative operator over the total space $\mathcal{A}\times\mathcal{A}_5$, and equivalently is the analytic form for extended BRST derivative operator, $\delta_{\emph{\emph{ex}}}:=\delta+\delta^*$, acting on elements of $C^\infty (\mathcal{A}\times\mathcal{A}_5)$.
\par \indent Note that using the same procedure, it is seen that for the ordinary geometric feature of BRST quantization, group indexed anti-ghosts and axial gauge fileds never appear in (\ref {3-1-8}) to form the exterior derivative operator. Therefore, for such cases, (\ref {3-1-8}) is replaced by;
\begin{eqnarray} \label {3-1-9}
\sum_{a,b,c,\mu}\int_M\{-\partial_\mu \omega^a(x)+C^{abc} A_\mu^b(x)\omega^c(x)\}(\frac{\delta}{\delta A_\mu^a (x)})~,
\end{eqnarray}
\par \noindent while the anti-ghost part is exclusively related to the axial extension;
\begin{eqnarray} \label {3-1-10}
\sum_{a,b,c,\mu}\int_M
\{C^{abc}B_\mu^b(x)\omega^{*c}(x)\}(\frac{\delta}{\delta A_\mu^a (x)})+\{-\partial_\mu \omega^{*a}(x)+C^{abc} A_\mu^b (x)\omega^{*c}(x)\}(\frac{\delta}{\delta B_\mu^a (x)})~.
\end{eqnarray}
\par \noindent Since each fiber of $\mathcal{A}\times\mathcal{A}_5$ is canonically diffeomorphic with $\tilde{\mathcal{G}}$, it is seen that the ghost part of (\ref {3-1-7}) defines the analytic form of BRST derivative. Specially, for any $\Gamma \in C^\infty (\mathcal{A}\times\mathcal{A}_5)$ one has,
\begin{eqnarray} \label {3-1-11}
\delta\Gamma=\sum_{a,b,c,\mu}\int_M
\{-\partial_\mu \omega^a(x)+C^{abc} A_\mu^b(x)\omega^c(x)\}(\frac{\delta\Gamma}{\delta A_\mu^a (x)})+\{C^{abc} B_\mu^b(x)\omega^c(x)\}(\frac{\delta\Gamma}{\delta B_\mu^a (x)})~.
\end{eqnarray}
\par \noindent Similarly it is seen that (\ref {3-1-10}) defines the analytic form of anti-BRST derivative operator. Thus, for any $\Gamma\in C^\infty (\mathcal{A}\times\mathcal{A}_5)$ we have;
\begin{eqnarray} \label {3-1-12}
\delta^*\Gamma=
\sum_{a,b,c,\mu}\int_M
\{C^{abc}B_\mu^b(x)\omega^{*c}(x)\}(\frac{\delta\Gamma}{\delta A_\mu^a (x)})+\{-\partial_\mu \omega^{*a}(x)+C^{abc} A_\mu^b (x)\omega^{*c}(x)\}(\frac{\delta\Gamma}{\delta B_\mu^a (x)})~.
\end{eqnarray}
\par ~

%%%%%%%%%%%%%%%%%%%%%%%%%%%%%%%%%%%%%%%%%%%

\par
\subsection{Modified Anomalies and Schwinger Terms}
\par Obviously, the forms of consistent anomalies and consistent Schwinger terms completely depend on the analytic form of BRST derivative operator via the descent equations. Thus, it is expected that when the algebra of ghost fields and eventually the BRST transformations get larger, then the consistent anomalies and Schwinger terms should be modified properly.
 \par By definition, gauge anomaly is the deviation of a second quantized Yang-Mills theory from its classical gauge symmetry. It is well-known that, anomalies cannot be canceled out properly by renormalization counter terms while in return, they ruin the renormalizability of the gauge theory [2, 34, 54]. In fact, classically the equations of motion of the Lagrangian density (\ref {1-1-3}) show that the vector and the axial currents, should obey the following conservation laws;
\begin{eqnarray} \label {3-2-1}
\begin{array}{c}
  D^a J(x):=\partial_\mu J^{a\mu} (x)+C^{abc}A_\mu^b(x) J^{c \mu)}(x)+C^{abc} B_\mu^b(x)J_5^{c\mu} (x)=0~, \\
  D_5^a J(x):=\partial_\mu J_5^{a\mu}(x)+C^{abc}A_\mu^b(x) J_5^{c\mu}(x)+C^abc B_\mu^b(x)J^{c\mu} (x)=0~.
\end{array}
\end{eqnarray}
\par On the other hand, the variation of the quantum action $W$ with respect to $A_\mu^a (x)$ (resp. $B_\mu^a (x)$) is $J^{a\mu} (x)$ (resp. $J_5^{a\mu}(x)$). Thus, from (\ref {3-1-11}) and (\ref {3-1-12}) we have;
\begin{eqnarray} \label {3-2-2}
\begin{array}{c}
  \sum_a\int_M\omega^a(x) D^aJ(x)=\delta W~,~~ \\
  \sum_a\int_M\omega^{*a}(x) D_5^aJ(x)=\delta^*W~.
\end{array}
\end{eqnarray}
\par \noindent Thus, the properties of $\delta$ and $\delta^*$ asserts that;
\begin{eqnarray} \label {3-2-3}
\begin{array}{c}
  \delta(\sum_a\int_M\omega^a(x) D^a J(x))=0~,~~~~~~~~~~~~~~~~~~~~~~~~~~~~~~~~~~~~~ \\
  \delta^*(\sum_a\int_M\omega^{*a}(x) D_5^a J(x))=0~,~~~~~~~~~~~~~~~~~~~~~~~~~~~~~~~~~~ \\
  \delta^*(\sum_a\int_M\omega^a(x) D^a J(x))+\delta(\sum_a\int_M\omega^{*a}(x) D_5^a J(x))=0~,
\end{array}
\end{eqnarray}
\par \noindent where the first equation is called the consistency condition [5, 22, 23] (with $B=0$). The second equation is somehow unfamiliar despite of its relation to axial currents. It is known that if a theory is anomalous, then both $\delta W$ and $\delta^* W$ participate in anomalous behavior simultaneously [4]. Thus, it is more convenient to consider $\delta_{\emph{\emph{ex}}}W=\delta W+\delta^* W$ as the anomalous term. Actually, the counter terms affect this description of anomalies such that a suitable choice of counter terms may lead to $\delta W=0$ as a desired result. But, cohomologically the extended BRST class of $\delta_{\emph{\emph{ex}}} W$ is unaffected with respect to counter terms. Indeed, the cohomology class of $\delta_{\emph{\emph{ex}}} W$ is independent of renormalization methods; in the other words, $\delta_{\emph{\emph{ex}}}W$ reveals the pathology structure of renormalizing the theory [23, 54]. Thus, extended BRST derivation produces a framework to study the anomalous behaviors of gauge theories.
\par \noindent Geometrically, using the Quillen supper connection [55] and the family index theory [56-57], it can be seen that $\Lambda:=\int_M\delta_{\emph{\emph{ex}}} W$ is the Cartan connection form of the principal $U(1)$-bundle due to the determinant line bundle $\widehat{\emph{\emph{DET}}}$, constructed from kernel of perturbed Dirac operators $D_{(A,B)}$ over $\mathcal{A}\times \mathcal{A}_5$, for extended gauge fields, $(A,B)\in \mathcal{A}\times \mathcal{A}_5$, as compact operators [37, 56, 58]. Any $U(1)$-gauge transformation causes the anomalous term $\Lambda$ to be added by a local term over $\mathcal{A}\times \mathcal{A}_5$, say $\lambda$. Moreover, this local term is an exact extended BRST form, i.e. $\lambda=\delta_{\emph{\emph{ex}}} w$, for $w$ an element of $C^\infty(\mathcal{A}\times \mathcal{A}_5)$, which never changes the cohomology class of $\Lambda$. Specially, such $\lambda$s produce exactly the whole set of possible counter terms appearing in renormalization the theory. Actually, (\ref {3-2-3}) shows that the connection of Quillen determinant bundle, restricted to the fibers of $\mathcal{A}\times \mathcal{A}_5$, is flat, i.e.; $\delta_{\emph{\emph{ex}}}\Lambda=0$, which is a geometric description of consistency conditions\footnote{By vanishing the curvature, i.e.; $\delta_{\emph{\emph{ex}}}\Lambda=\delta \Lambda +\delta^* \Lambda=0$, we basically focus on the whole three equations of (\ref {3-2-3}) as extended consistency conditions. In fact, in the extended BRST formalism, the standard consistency condition, $\delta W=0$, is simultaneously replaced with three equations of (\ref {3-2-3})} [37].\\
\par \indent In non-commutative geometry, it is known that this anomalous behavior arises because $(A+B)D^{-1}$, for free Dirac operator $D$, is not a trace class operator in general [59, 60]. Actually, $(A+B)D^{-1}$ belongs to the Schatten class $\ell^p (L^2 (\mathcal{S}(M)\otimes E))$ for $p>\dim M$, which causes the quantum action $W$ not to be well defined over $\mathcal{A}\times \mathcal{A}_5$. To see this recall that;
\begin{eqnarray} \label {3-2-4}
\frac{\int_{\psi,\bar{\psi}}e^{iS(\psi,\bar{\psi},A,B)}}{\int_{\psi,\bar{\psi}}e^{iS(\psi,\bar{\psi})}}=\exp(~\sum_{n=1}^\infty\frac{(-1)^{n+1}}{n}~ Tr\{((A+B) D^{-1})^n\}~)~.
\end{eqnarray}
\par \noindent But for small enough integers $n$, $n\leq\dim M$, $((A+B) D^{-1})^n$s are not trace class operators. Then, one needs to modify the trace. This can be considered as the regularization method in the setting of operator theory. Such fascinating feature of regularization takes place by using the Dixmier trace and zeta Riemann function [59-61]. This causes the quantum action $W$ to vary through the fibers. Indeed, $W$ is not even a single valued function over $\mathcal{A}\times\mathcal{A}_5$. Therefore, $\delta_{\emph{\emph{ex}}} W$ belongs to a nontrivial deRham (extended BRST) cohomology class of  $\tilde{\mathcal{G}}$ [23, 59, 60]. As an intuitive example for this situation one can consider the angle function $\theta$ over $S^1$. $\theta$ is not a continuous function but its exterior derivative $\emph{\emph{d}}\theta$, forms a nontrivial cohomology class of $H_{\emph{\emph{deR}}}^1 (S^1,\mathbb{R})$. Thus, integrating the pull back of $\delta_{\emph{\emph{ex}}} W$ through a smooth map $g:S^1\rightarrow \tilde{\mathcal{G}}$ over $S^1$ measures the anomalous behavior of the theory [59, 60, 62].  In this way, one computes an element of the holonomy group of the Quillen determinant bundle $\widehat{\emph{\emph{DET}}}$ over $\tilde{\mathcal{G}}$. As it was stated above, this bundle is equipped with a flat connection which leads to a projective group homomorphism of $\phi:\pi_1(\tilde{\mathcal{G}})=\pi_{1+\dim M} (\tilde{G})\twoheadrightarrow Hol(\widehat{\emph{\emph{DET}}})$. Actually, since this connection preserves the metric of $\widehat{\emph{\emph{DET}}}$, then $Hol(\widehat{\emph{\emph{DET}}})\subseteq U(1)$ [56-58]. But since $Z(A+B)$, the partition function of an extended gauge theory with extended gauge field $(A,B)$, is single valued over $\mathcal{A}\times\mathcal{A}_5$, then $Hol(\widehat{\emph{\emph{DET}}})=\{1\}$. Therefore, one concludes that $\int_{S^1} g^*(\Lambda)=2\pi m$ for $m$ the winding number of the phase of $\det D_{(A,B)\lhd g}$ around the loop $g:S^1\rightarrow \tilde{\mathcal{G}}$ [62]. This can be computed in terms of periodic cyclic cohomology, using the Chern-Connes character and the local index formula [59-61].\\
\par  To extract the modified consistent anomaly according to the Stora-Zumino procedure [18, 19], one should use $\emph{\emph{d}}$, $\delta$ and $\delta^*$ alternatively to provide a generalized formulation of descent equations. Initially, the Bianchi identity asserts that;
\begin{eqnarray} \label {3-2-5}
\begin{array}{c}
  \delta R=-i\omega R+iR\omega~,~~~~~~~~~~~~~~~ \\
  \delta^* R=-i\omega^* R+iR\omega^*~,~~~~~~~~~~~ \\
  \emph{\emph{d}}R=i(A+B)R-iR(A+B)~,
\end{array}
\end{eqnarray}
\par \noindent where $R=-i\emph{\emph{d}}A-A^2$ is the curvature\footnote{We should emphasize again that the gauge fields and consequently the curvature tensors are here considered in representation of which they act on the spinors, as they appear in the quantum action}. Thus, (\ref {3-2-5}) implies that $Tr\{R^{n+1}\}$ is a deRham, BRST, and anti-BRST closed form. Set $M=\mathbb{R}^{2n}$ and consider $Tr\{R^{n+1}\}$ as a $2n+2$-form over $\mathbb{R}^{2n+2}$. Thus, the Poincare lemma leads to;
\begin{eqnarray} \label {3-2-6}
\begin{array}{c}
  Tr\{R^{n+1}\}=\emph{\emph{d}}\Omega_{2n+1}^{0,0}~, \\
  \begin{array}{cc}
    \delta\Omega_{2n+1}^{0,0}=\emph{\emph{d}}\Omega_{2n}^{1,0}~,~~~~~~~~~~~~~~~~~~~~~~~~~~~~~~ & \delta^* \Omega_{2n+1}^{0,0}=\emph{\emph{d}}\Omega_{2n}^{0,1}~, \\
    \delta\Omega_{2n}^{1,0}=\emph{\emph{d}}\Omega_{2n-1}^{2,0}~,~~~~~~~~~~~~~~~~~~~~~~~~~~~~~~ & \delta^* \Omega_{2n}^{0,1}=\emph{\emph{d}}\Omega_{2n-1}^{0,2},
  \end{array}
 \\
  \delta^* \delta\Omega_{2n+1}^{0,0}=\emph{\emph{d}}\Omega_{2n}^{1,1}~,
\end{array}
\end{eqnarray}
\par \noindent where $\Omega_i^{j,k}$ is a deRham differential $i$-form with ghost number of $j-k$, while $i+j+k=2n+1$. Actually, $\Omega_i^{j,k}$ is simultaneously a differential $j$-form over $\mathcal{A}$ and a differential $k$-form over $\mathcal{A}_5$. On the other hand, it can easily be shown that $\Omega_{2n}^{1,1}=\delta \Omega_{2n}^{0,1}=-\delta^*\Omega_{2n}^{1,0}+\emph{\emph{d}}\Omega_{2n-1}^{1,1}$. Indeed, $\delta\Omega_{2n}^{0,1}$ and $-\delta^* \Omega_{2n}^{1,0}$ differ in an exact differential form. This fact plays an important role in calculating the consistent Schwinger term.
\par \indent From (\ref {3-2-6}) it is seen that;
\begin{eqnarray} \label {3-2-7}
\delta_{\emph{\emph{ex}}}(\Omega_{2n}^{1,0}+\Omega_{2n}^{0,1})=\emph{\emph{d}}(\Omega_{2n-1}^{2,0}+\Omega_{2n-1}^{1,1}+\Omega_{2n-1}^{0,2})~,
\end{eqnarray}
\par \noindent and hence;
\begin{eqnarray} \label {3-2-8}
\delta_{\emph{\emph{ex}}}\int_{\mathbb{R}^{2n}} (\Omega_{2n}^{1,0}+\Omega_{2n}^{0,1})=0~.
\end{eqnarray}
\par Therefore, up to a factor $(\Omega_{2n}^{1,0}+\Omega_{2n}^{0,1})(x)$ can be considered as the modified non-integrated consistent anomaly; in the other words, $\int_{\mathbb{R}^{2n}} (\Omega_{2n}^{1,0}+\Omega_{2n}^{0,1})$  is a candidate for $\delta_{\emph{\emph{ex}}}W=\delta W+\delta^*W$. Here we should emphasize that despite of the effect of counter terms to anomalous descriptions, which makes us apply the term of "candidate" for consistent anomaly and consistent Schwinger term, the extended BRST cohomology class of $\Omega_{2n+1-i-j}^{i,j}$ is exactly unaffected by renormalization methods. In fact, in the following we precisely prove that the extended BRST cohomology class of $\int_{\mathbb{R}^{2n}} (\Omega_{2n}^{1,0}+\Omega_{2n}^{0,1})$ coincides exactly with that of $\delta_{\emph{\emph{ex}}}W$.\\
\par \noindent A direct calculation shows that when $n=2$ then;
\begin{eqnarray} \label {3-2-9}
\Omega_4^{1,0}+\Omega_4^{0,1}=Tr\{\emph{\emph{d}}(\omega+\omega^*)(~i(A+B)\emph{\emph{d}}(A+B)+\frac{1}{2}(A+B)^3~)\}
\end{eqnarray}
which is the modified consistent anomaly up to a factor of $c_2=\frac{1}{24\pi^2}$. Indeed, if one sets $B=0$ the resulting form is equal to the well-known consistent anomaly\footnote{Actually, by setting $B=0$ anti-ghost fields will be killed automatically after taking the trace.} [23, 54]. On the other hand, ghost number counting leads to;
\begin{eqnarray} \label {3-2-10}
c_2 \Omega_4^{1,0}
=\frac{1}{24\pi^2} ~ Tr\{\emph{\emph{d}}\omega(~i(A+B)\emph{\emph{d}}(A+B)+\frac{1}{2}(A+B)^3~)\}~,
\end{eqnarray}
\par \noindent which we called "\emph{ghost consistent anomaly}". Moreover, the remained term;
\begin{eqnarray} \label {3-2-11}
c_2 \Omega_4^{0,1}
=\frac{1}{24\pi^2} ~ Tr\{\emph{\emph{d}}\omega^*(~i(A+B)\emph{\emph{d}}(A+B)+\frac{1}{2}(A+B)^3~)\}~,
\end{eqnarray}
\par \noindent is also referred to as "\emph{anti-ghost consistent anomaly}".\\
\par \indent If $W =\int_{\mathbb{R}^{2n}}w$, for a non-local form $w$, then $\delta_{\emph{\emph{ex}}}W=c_n \int_{\mathbb{R}^{2n}}(\Omega_{2n}^{1,0}+\Omega_{2n}^{0,1})$  implies that;
\begin{eqnarray} \label {3-2-12}
\begin{array}{c}
  \emph{\emph{d}}w=-c_n \Omega_{2n+1}^{0,0}+K~, \\
  \emph{\emph{d}}w=-c_n \Omega_{2n+1}^{0,0}+K'~,
\end{array}
\end{eqnarray}
\par \noindent for a BRST (resp. anti-BRST) closed form $K$ (resp. $K'$). Thus, the added term $K$ (or $K'$) must be simultaneously a BRST and an anti-BRST closed form. This undetermined BRST/anti-BRST (extended BRST) closed form can be compared with the gauge fixing term in the Faddeev-Popov quantization method [12, 54]. It asserts that the quantum Lagrangian density is simultaneously invariant under BRST and anti-BRST transformations, the fact of which was studied in [17] with details.\\
\par \indent Moreover, continuing the sequence of equations (\ref {3-2-6}), results in;
\begin{eqnarray} \label {3-2-13}
\delta\Omega_{2n-1}^{2,0}=\emph{\emph{d}}\Omega_{2n-2}^{3,0}~~~~~~~    ,   ~~~~~~~  \delta^* \Omega_{2n-1}^{0,2}=\emph{\emph{d}}\Omega_{2n-2}^{0,3}~ .
\end{eqnarray}
\par \noindent Indeed, fallowing the standard descent equations for $\emph{\emph{d}}$ and $\delta_{\emph{\emph{ex}}}=\delta+\delta^*$ yields the following equality,
\begin{eqnarray} \label {3-2-14}
\delta_{\emph{\emph{ex}}}(\Omega_{2n-1}^{2,0}+\Omega_{2n-1}^{0,2}+\Omega_{2n-1}^{1,1})
=\emph{\emph{d}}(\Omega_{2n-2}^{3,0}+\Omega_{2n-2}^{2,1}+\Omega_{2n-2}^{1,2}+\Omega_{2n-2}^{0,3})~.
\end{eqnarray}
\par \noindent Ghost number counting implies that,
\begin{eqnarray} \label {3-2-15}
\begin{array}{c}
  \delta\Omega_{2n-1}^{2,0}=\emph{\emph{d}}\Omega_{2n-2}^{3,0}~, \\
  \delta^*\Omega_{2n-1}^{0,2}=\emph{\emph{d}}\Omega_{2n-2}^{0,3}~,
\end{array}
\end{eqnarray}
\par \noindent and,
\begin{eqnarray} \label {3-2-16}
\begin{array}{c}
  \delta^* \Omega_{2n-1}^{2,0}+\delta\Omega_{2n-1}^{1,1}=\emph{\emph{d}}\Omega_{2n-2}^{2,1}~  , \\
  \delta\Omega_{2n-1}^{0,2}+\delta^*\Omega_{2n-1}^{1,1}=\emph{\emph{d}}\Omega_{2n-2}^{1,2}~.
\end{array}
\end{eqnarray}
\par \noindent Therefore,
\begin{eqnarray} \label {3-2-17}
\Omega_{2n-1}^{2,0}+\Omega_{2n-1}^{0,2}+\Omega_{2n-1}^{1,1}~  ,
\end{eqnarray}
\par \noindent is a candidate for the modified consistent Schwinger term up to a factor $c_n$. For $n=2$, the modified consistent Schwinger term is given by;
\begin{eqnarray} \label {3-2-18}
c_2 (\Omega_3^{2,0}+\Omega_3^{0,2}+\Omega_3^{1,1})
=\frac{1}{24\pi^2} ~ Tr\{(\emph{\emph{d}}(\omega+\omega^*))^2 (A+B)\}~.
\end{eqnarray}
\par \noindent where;
\begin{eqnarray} \label {3-2-19}
\begin{array}{c}
  c_2 \Omega_3^{2,0}=\frac{1}{24\pi^2} Tr\{(\emph{\emph{d}}\omega)^2 A\}~,~~~~~~~~~~~~~~~ \\
  c_2 \Omega_3^{0,2}=\frac{1}{24\pi^2} Tr\{(\emph{\emph{d}}\omega^*)^2 A\}~,~~~~~~~~~~~~~~ \\
  c_2 \Omega_3^{1,1}=\frac{1}{24\pi^2} Tr\{(\emph{\emph{d}}\omega \emph{\emph{d}}\omega^*+\emph{\emph{d}}\omega^*\emph{\emph{d}}\omega)B\}~.
\end{array}
\end{eqnarray}
\par \noindent From now on we refer to these differential forms with "\emph{ghost/ghost}", "\emph{anti-ghost/anti-ghost}" and "\emph{ghost/anti-ghost}" consistent Schwinger term, respectively. Note that the \emph{ghost consistent anomaly} and the \emph{ghost/ghost consistent Schwinger term} are related by an ordinary descent equation of $\emph{\emph{d}}$ and $\delta$. It is also the case for \emph{anti-ghost consistent anomaly} and \emph{anti-ghost/anti-ghost consistent Schwinger term} with replacing $\delta$ by $\delta^*$. More generally, the extended descent equation will give rise to a lattice diagram which commutes up to exact deRham forms. This lattice, produces a bi-complex for horizontal maps of $(\delta,\emph{\emph{d}})$, and vertical ones with $(\delta^*,\emph{\emph{d}})$ as;\\
\begin{eqnarray} \label {3-2-20}
\begin{array}{ccccccccc}
  \Omega_{2n+1}^{0,0} & \leftrightarrow
 & \Omega_{2n}^{1,0} & \leftrightarrow  & \Omega_{2n-1}^{2,0} & \leftrightarrow  & \Omega_{2n-2}^{3,0} & \leftrightarrow & \ldots \\
  \updownarrow &    & \updownarrow &    & \updownarrow &    & \updownarrow &    &   \\
  \Omega_{2n}^{0,1} & \leftrightarrow & \Omega_{2n-1}^{1,1} & \leftrightarrow & \Omega_{2n-2}^{2,1} & \leftrightarrow & \Omega_{2n-3}^{3,1} & \leftrightarrow & \ldots \\
  \updownarrow &    & \updownarrow &    & \updownarrow &    & \updownarrow &    &   \\
  \Omega_{2n-1}^{0,2} & \leftrightarrow & \Omega_{2n-2}^{1,2} & \leftrightarrow & \Omega_{2n-3}^{2,2} & \leftrightarrow & \Omega_{2n-4}^{3,2} & \leftrightarrow & \ldots \\
  \updownarrow &    & \updownarrow &    & \updownarrow &    & \updownarrow &    &   \\
  \Omega_{2n-2}^{0,3} & \leftrightarrow & \Omega_{2n-3}^{1,3} & \leftrightarrow & \Omega_{2n-4}^{2,3} & \leftrightarrow & \Omega_{2n-5}^{3,3} & \leftrightarrow & \ldots \\
  \updownarrow &    & \updownarrow &    & \updownarrow &   & \updownarrow &    &   \\
  \vdots &   & \vdots &   & \vdots &   & \vdots &   & \ddots
\end{array}
\end{eqnarray}\\
\par In the diagram of (\ref {3-2-20}) the differential forms $\Omega_{2n+2-i-j/2n+1-i-j}^{i,j}$ at the inner vertices are considered as $\Omega_{2n+2-i-j}^{i,j}$ (resp. $\Omega_{2n+1-i-j}^{i,j}$) for the range (resp. domain) of the relevant incoming (resp. outgoing) arrows. Finally, it can be seen that the co-diagonal elements of (\ref {3-2-20}) (the elements of $\Omega_{2n+1-i-j}^{i,j}$ with the same number of $i+j$) produce modified consistent anomalies, modified consistent Schwinger terms and so on. More precisely, the extended formalism of descent equations is given for the co-diagonals of (\ref {3-2-20}). Note that the top row and the left column of (\ref {3-2-20}) show respectively the ordinary deRham/BRST and deRham/anti-BRST descent equations as was mentioned above.\\
\par \indent Here for the last description we should precisely prove that the extended BRST cohomology class of $\int_{\mathbb{R}^{2n}} (\Omega_{2n}^{1,0}+\Omega_{2n}^{0,1})$ coincides exactly with that of $\delta_{\emph{\emph{ex}}}W$. To show this note that we are essentially looking for a deRham differential $2n$-form with (anti-) ghost number equal to one, as the non-integrated consistent anomaly. Physically we need an observable with mass dimension $2n$ and (anti-) ghost number equal to one. This differential form is basically a part of a differential $2n+1$-form over $\tilde{\mathcal{G}}\times M$, which can be decomposed as;
\begin{eqnarray} \label {BRST cohomology}
\Xi_{2n+1}=\sum_i^{2n+1} (-1)^i~\Xi_{2n+1-i}^i=\Xi_{2n+1}^0-\Xi_{2n}^1+\Xi_{2n-1}^2-...~,
\end{eqnarray}
\par \noindent where here $i$ is the extended ghost number, i.e. the sum of ghost and anti-ghost numbers. Moreover, in (\ref {BRST cohomology}) we let the differential forms on $M$ to get degrees higher than $2n$, the dimension of $M$. However, if $\Xi_{2n+1}$ defines a deRham cohomology class then, $(\delta_{\emph{\emph{ex}}}+\emph{\emph{d}})\Xi_{2n+1}$ should vanish. Therefore, by extended ghost number counting we have;
\begin{eqnarray}
\begin{array}{c}
  \emph{\emph{d}}\Xi_{2n+1}^0=0~,~~ \\
  \delta_{\emph{\emph{ex}}}\Xi_{2n+1}^0=\emph{\emph{d}}\Xi_{2n}^1~, \\
  ~~~~~~\delta_{\emph{\emph{ex}}}\Xi_{2n}^1=\emph{\emph{d}}\Xi_{2n-1}^2~, \\
  ~~~~\vdots
\end{array}
\end{eqnarray}
\par \noindent which except the first equation produces properly the extended descent equations of (\ref {3-2-6}). As stated above, topologically we demand the pull back of $\int_M\Xi_{2n}^1$ through any smooth $g:S^1\hookrightarrow \tilde{\mathcal{G}}$ defines an element of $H^1(S^1,\mathbb{Z})$. Applying the index theorem and the Poincare lemma for the case of $M=\mathbb{R}^{2n}$ we deduce that $\emph{\emph{d}}\Xi_{2n+1}^0$ doesn't vanish, but in return, as the only choice, it coincides with $c_nTr\{R^{n+1}\}$ with null cohomology class, and with specific topological constant $c_n$, as we discussed above. Now, applying the dimension of $M$, we get a topological cohomology class with;
\begin{eqnarray} \label {BRST cohomology2}
\Omega_{2n+1}=\sum_i^{2n+1} (-1)^{i+1}~\Omega_{2n+1-i}^i=\Omega_{2n}^1-\Omega_{2n-1}^2+...~,
\end{eqnarray}
\par \noindent deduced thoroughly from $(n+1)$th Chern character $c_nTr\{R^{n+1}\}$ which essentially vanishes. On the other hand, it is clear that by adding any extended BRST exact form to $\Omega_{2n}^1$, the cohomology class of $\Omega_{2n+1}$ doesn't change. This basically rejects the effect of renormalization methods on the BRST class of anomaly terms. This precisely proves our claim. 
%%%%%%%%%%%%%%%%%%%%%%%%%%%%%%%%%%%%%%%%%%

\section{Conclusions}
%%%%%%%%%%%%%%%%%%%%%%%%%%%%%%%%%%%%%
\par In this article it was shown that in gauge theories, anti-BRST invariance is the quantized version of local axial symmetry. Basically, the axially extension of a gauge theory enlarges the gauge group and consequently the algebra of ghost fields. Initially, it was shown that the anti-BRST derivation and infinitesimal axial transformations are mutually dual in the sense of differential geometric objects. Moreover, an elaborate geometric description for anti-ghost and anti-BRST transformation was given by means of differential objects over the space of connections of the gauge theory principal bundle. Finally, a collection of extended descent equations was formulated by using BRST and anti-BRST derivatives alternatively. This results led to a modified version for consistent anomalies and consistent Schwinger terms.

%%%%%%%%%%%%%%%%%%%%%%%%%%%%%%%%%%%%%%%%%%%%%%%%
\section{Acknowledgments}
%%%%%%%%%%%%%%%%%%%%%%%%%%%%%%%%%%%%%%%%%%%%%%%%
\par The author expresses his gratitude to F. Ardalan, R. Bertlmann, A. Davody, D. Freed, M. Khalkhali, D. Perrot and A. Shafiei Deh Abad for their use-full comments. However my special thanks go to M. M. Sheikh-Jabbari for his elegant remarks and pedagogical discussions. Also the author should confess that this article owes most of its appearance to S. Ziaee for many reasons. Finally, I should recall that most parts of this research was done during my PhD period in Sharif University of Technology.

%%%%%%%%%%%%%%%%%%%%%%%%%%%%%%%%%%%%%%%%%%%%%%%%%%%%%%%%%%%%%%%%%%
%\section{Appendices}
%%%%%%%%%%%%%%%%%%%%%%%%%%%%%%%%%%%%%%%%%%%%%%%%%%%%%%%%%%%%%%%%%%%
%\begin{appendix}\setcounter{equation}{0}\noindent
%\section{}\

%%%%%%%%%%%%%%%%%%%%%%%%%%%%%%%%%%%%%%%%%

\end{document}